\documentclass[epjST]{svjour}
\usepackage{graphics}
\usepackage{arydshln}
\begin{document}

\title{Electrodynamic response of Ba(Fe$_{1-x}$Rh$_x$)$_2$As$_2$ across the s$_{\pm}$ to s$_{++}$ order parameter transition}

\author{D. Torsello\inst{1}$^,$\inst{2}\fnmsep\thanks{\email{daniele.torsello@polito.it}} \and R. Gerbaldo\inst{1}$^,$\inst{2} \and L. Gozzelino\inst{1}$^,$\inst{2} \and M. A. Tanatar\inst{3}$^,$\inst{4} \and R. Prozorov\inst{3}$^,$\inst{4} \and P. C. Canfield\inst{3}$^,$\inst{4} \and G. Ghigo\inst{1}$^,$\inst{2}}
\institute{Politecnico di Torino, Department of Applied Science and Technology, Torino 10129, Italy \and Istituto Nazionale di Fisica Nucleare, Sezione di Torino, Torino 10125, Italy \and Ames Laboratory, US Department of Energy, Ames, Iowa 50011, USA \and Department of Physics \& Astronomy, Iowa State University, Ames, Iowa 50011, USA}
\abstract{
Most iron-based superconductors are characterized by the s$_{\pm}$ symmetry of their order parameter, and are expected to go through a transition to the s$_{++}$ state if enough disorder is introduced. We previously reported the observation of this transition in Ba(Fe$_{1-x}$Rh$_x$)$_2$As$_2$ through a study of the disorder dependence of the critical temperature and low-temperature London penetration depth. In this paper we report on the analysis of the electrodynamic response of the same sample across the transition and we identify peculiarities in the behaviour of the surface resistance and normal conductivity, that can be considered as traces of the transition itself. 
} 
\maketitle
\section{Introduction}
\label{intro}
Most iron-based superconductors (IBS), and in particular those of the 122 family, such as doped BaFe$_2$As$_2$, are characterized by a fully gapped pairing state with sign changing over different Fermi sheets: the s$_{\pm}$ state \cite{Chubukov2008}. These systems present multiple bands that cross the Fermi level and on which superconducting s-wave gaps open \cite{Stanev2008}. The relative sign of these gaps is opposite on hole and electron bands and coupling between them is provided mainly by antiferromagnetic spin fluctuations \cite{Mazin2008}.\\
Although direct experimental observations of the symmetry of the pairing state are extremely difficult to achieve, the s$_{\pm}$ state can be indisputably assigned through the identification of the disorder-driven transition to the s$_{++}$  sign-preserving state predicted by Efremov \textit{et} \textit{al.} \cite{Efremov2011}. In any multi-band superconductor, disorder causes the scattering of particles between different bands, that in turn drives the values of the gaps towards a common value \cite{Wang2013}. If in the pristine system the gaps have different sign, it is necessary for the system to reach a state in which all the gaps have the same sign in order for their values to converge: this is the s$_{\pm}$ to s$_{++}$ transition.\\
\sloppy This behaviour was first observed by Schilling \textit{et} \textit{al.} \cite{Schilling2016} in thin films of Ba(Fe$_{1-x}$Co$_x$)$_2$As$_2$ by extracting the value of the small gap from the coherence peak in the measured optical conductivity at increasing levels of disorder introduced via electron irradiation. In our previous work on this topic \cite{Ghigo2018prl}, we identified the  s$_{\pm}$ to s$_{++}$ transition in proton irradiated Ba(Fe$_{1-x}$Rh$_x$)$_2$As$_2$ single crystals by observing one of its predicted hallmarks: a drop in the low-temperature value of the penetration depth. The experimental observation was validated by two-bands Eliashberg calculations that reproduced the experimental critical temperatures and superfluid densities, with the transition visible in the change of sign of the smallest gap at the expected disorder level. Those penetration depth and critical temperature measurements were carried out by means of a microwave resonator technique that also yields the normal conductivity $\sigma_n$ and surface impedance $Z_s$ as a function of temperature, below and above the critical temperature $T_c$ \cite{Ghigo2016ieee}.\\ 
In this work we analyze the behaviour of $\sigma_n$ and $Z_s$ across the s$_{\pm}$ to s$_{++}$ transition and we identify peculiarities that can be considered as traces of the transition itself.
\section{Experiment and results}
\label{sec:1}

\noindent \sloppy The sample under investigation is the same optimally doped single crystal of  Ba(Fe$_{1-x}$Rh$_x$)$_2$As$_2$ studied in our previous work \cite{Ghigo2018prl}. It was grown out of self flux using conventional high-temperature solution growth techniques \cite{Ni2008,Ni2009}, resulting in a doping level $x=0.068$ (determined by wavelength dispersive X-ray spectroscopy  \cite{Hodovanets2012}) and in a pristine critical temperature $T_c=23.5$ K. The critical temperature considered here is the temperature at which the penetration depth diverges, as reported in \cite{Ghigo2018prl}. Disorder was introduced in the system in the form of point-like defects and small cascades via successive 3.5-MeV proton irradiation sessions. The amount of induced disorder was estimated as displacements per atom (d.p.a.) via Monte Carlo simulations with PHITS \cite{phits} and SRIM \cite{srim} codes.

\begin{figure}[h!]
\begin{center}
\resizebox{0.75\columnwidth}{!}{%
  \includegraphics{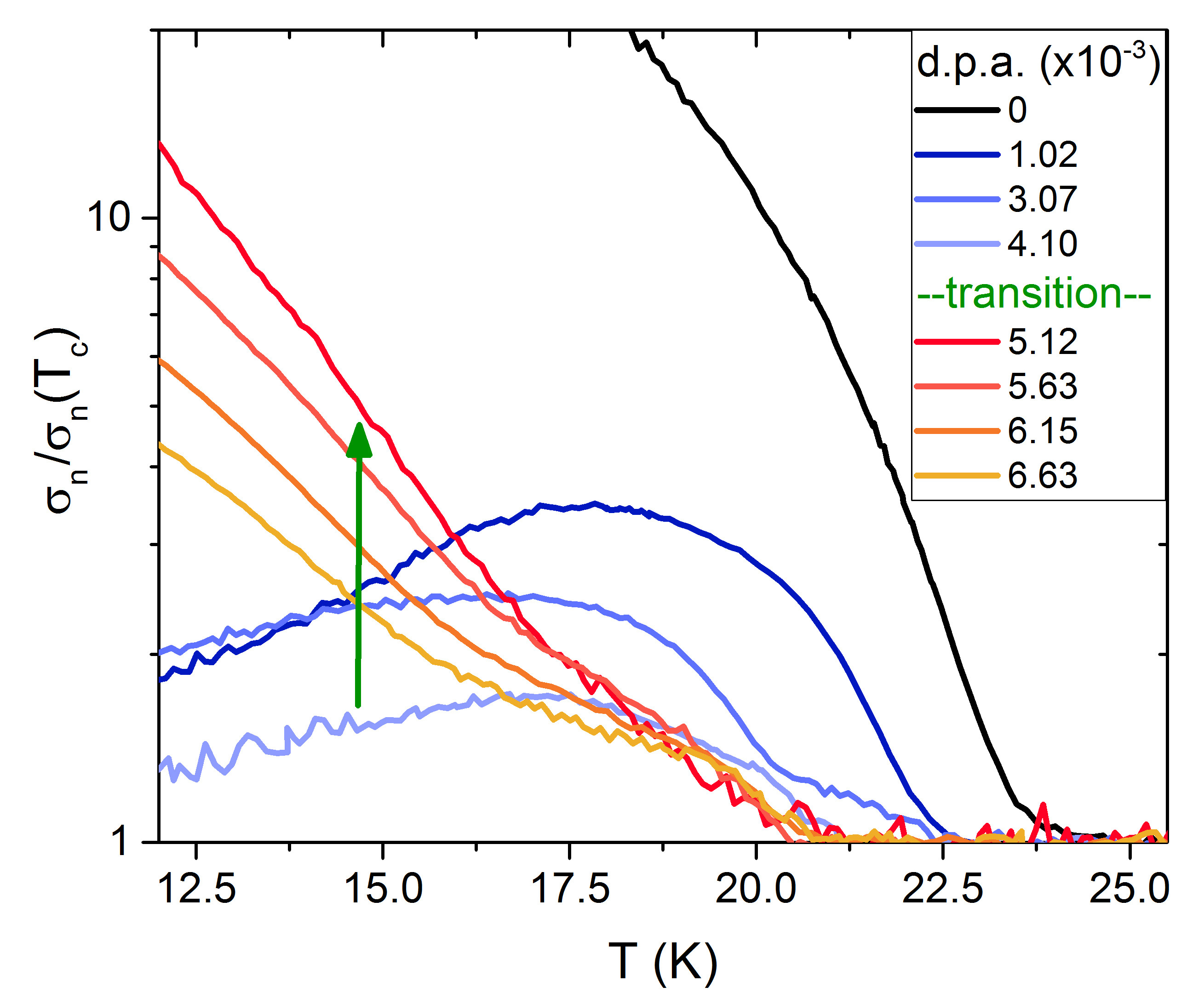} }
\caption{Normalized quasiparticle conductivity vs temperature curves for increasing levels of disorder (d.p.a.). The s$_{\pm}$ to s$_{++}$ transition takes place between d.p.a. = 4.10$\times10^{-3}$ and 5.12$\times10^{-3}$, and is indicated by the green arrow.}
\label{fig:cond}    
\end{center}  
\end{figure}
\noindent Superconducting properties were characterized in the pristine state and after each irradiation session using a microwave resonator technique, already applied to IBS with different doping species, pristine \cite{Ghigo2017prb} and irradiated \cite{Ghigo2017scirep}. This technique consists in measuring the perturbations induced to the resonance of an YBa$_2$Cu$_3$O$_{7-x}$ coplanar waveguide resonator by coupling a small crystal to it. From the modifications in the resonance frequency and quality factor, and after a calibration procedure (explained in details in \cite{Ghigo2018sust} and \cite{Ghigo2017prb}), the London penetration depth, normal conductivity and surface impedance as a function of temperature can be obtained.\\
Penetration depth data was used in 
\cite{Ghigo2018prl} to identify the s$_{\pm}$ to s$_{++}$ transition. In the following, normal conductivity and surface impedance are analyzed.\\
Figure \ref{fig:cond} shows the temperature dependence of the normal conductivity $\sigma_n$, normalized to its value at $T_c$, for all the levels of disorder analyzed. The sample in the pristine state shows a sharp increase of $\sigma_n$ below the critical temperature. When disorder is introduced, but the s$_{\pm}$ to s$_{++}$ transition has not yet occurred, the curves have a qualitatively different shape: the conductivity at low temperature is strongly suppressed and shows a clear and large peak at approximately 17 K. Surprisingly, when the s$_{++}$ state is reached, the conductivity curves become very similar to that of the pristine state, although with values that decrease with increasing disorder.

\begin{figure}[h!]
\begin{center}
\resizebox{0.95\columnwidth}{!}{%
  \includegraphics{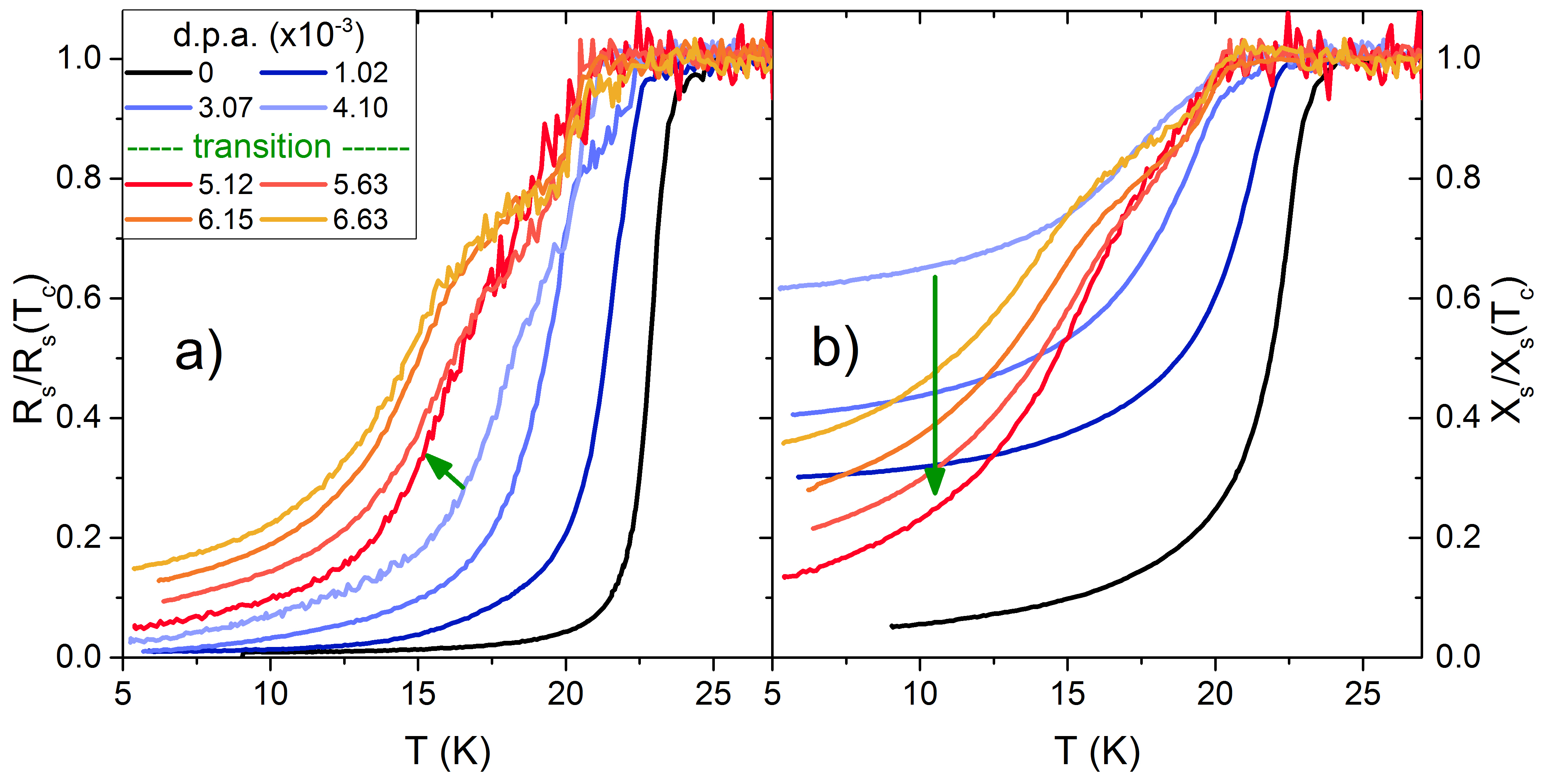} }
\caption{a) Normalized surface resistance ($R_s$) and b) reactance ($X_s$) vs temperature curves for increasing levels of disorder (d.p.a.).  The s$_{\pm}$ to s$_{++}$ transition takes place between d.p.a. = 4.10$\times10^{-3}$ and 5.12$\times10^{-3}$, and is indicated by the green arrows.}
\label{fig:Zs}   
\end{center}   
\end{figure}
\noindent The surface impedance $Z_s=R_s+iX_s$ as a function of temperature is shown in Figure \ref{fig:Zs} for the pristine sample and after each irradiation dose. Both $R_s$ and $X_s$ are normalized to their values at $T_c$, above which $R_s=X_s$ as expected in the classical skin effect regime. When disorder increases, the real and imaginary parts of the impedance behave differently: at low temperature the normalized surface resistance shows a monotonic increase, whereas low temperature value of the surface reactance drops at the transition (see Table \ref{tab:1} for quantitative values). Moreover, in the s$_{++}$ state $X_s$ also develops a shoulder near $T=15$ K.\\

\begin{table}[h]
\begin{center}
\caption{Normalized surface impedance values across the s$_{\pm}$ to s$_{++}$ transition.}
\label{tab:1}
\begin{tabular}{ccc}
\hline\noalign{\smallskip}
d.p.a.($\times10^{-3}$) & $R_s(10K)/R_s(T_c)$ & $X_s(10K)/X_s(T_c)$  \\
\noalign{\smallskip}\hline\noalign{\smallskip}
\centering 0    & 0.01 & 0.06 \\
1.02 & 0.01 & 0.32 \\
3.07 & 0.03 & 0.44 \\
4.10 & 0.07 & 0.65 \\
\noalign{\smallskip}\hdashline\noalign{\smallskip}
5.12 & 0.10 & 0.23 \\
5.63 & 0.14 & 0.30 \\
6.15 & 0.19 & 0.37 \\
6.63 & 0.22 & 0.46 \\
\noalign{\smallskip}\hline
\end{tabular}
\end{center}
\end{table}
\section{Discussion and conclusions}
\label{sec:2}
We report on measurements of normal conductivity and surface impedance across the s$_{\pm}$ to s$_{++}$ transition in proton irradiated Ba(Fe$_{1-x}$Rh$_x$)$_2$As$_2$ that show three peculiar signs of the transition itself:
\begin{itemize}
\item the normal conductivity after the transition recovers the monotonous trend with temperature that characterized the pristine state.
\item the low temperature value of the surface reactance drops at the transition, whereas that of the surface resistance increases monotonically with disorder.
\item the surface impedance versus temperature curves develop a shoulder below $T_c$ after the transition.
\end{itemize}

\noindent It is important to note that the normal conductivity (measured at a frequency of 8 GHz) is not the same quantity on which Schilling \textit{et} \textit{al.}\cite{Schilling2016} based their identification of the s$_{\pm}$ to s$_{++}$ transition. They measured the optical conductivity at zero frequency and observed, at all levels of disorder, the coherence peak that is centered at a temperature related to the width of the smallest gap. Instead, in our measurements we find a peak only for the irradiated system in the s$_{\pm}$ state. It could be originated either by coherence effects or by the competition between increasing quasiparticle scattering time and decreasing quasiparticle density, and it could be influenced and eventually masked also by other frequency-related effects \cite{Barannik2013,Takahashi2011}.\\
Regarding surface impedance, the hallmarks of the transition we identified can be observed with other microwave resonator techniques that directly give the surface resistance and reactance from the measured resonance frequency and quality factor. Moreover, the combination of the monotonic increase of resistance with the drop of reactance at the s$_{\pm}$ to s$_{++}$ transition could be investigated for superconducting applications in which a modulated reactance is needed with almost-constant resistance \cite{Panaretos}.

\section{Acknowledgements}
\label{Acknowledgements}
This research activity was perfomed in the framework of the INFN-Politecnico di Torino M.E.S.H. Research Agreement, and also supported by the Coordinated Research Project F11020 of the International Atomic Energy Agency (IAEA). The authors thank the INFN-LNL staff for the help with irradiation experiments.


\end{document}